\documentstyle[aps,epsf,preprint]{revtex}
\tightenlines
\begin{document}
\draft
\title{The new analysis of the KLOE data on the $\phi\to\eta\pi^0\gamma$ decay}
\author
{N.N. Achasov
\thanks{achasov@math.nsc.ru}
\ and A.V. Kiselev
\thanks{kiselev@math.nsc.ru}}

\address{
   Laboratory of Theoretical Physics,
 Sobolev Institute for Mathematics, Novosibirsk, 630090}

  \date{\today}
\maketitle $ $\\[2mm]

\begin{abstract}
In this paper we present the fit to the recent high-statistical
KLOE data on the $\phi\to\eta\pi^0\gamma$ decay. This decay mainly
goes through the $a_0\gamma$ intermediate state. The obtained
results differ from those of the previous fits: data prefer a high
$a_0$ mass and a considerably large $a_0$ coupling to the
$K\bar{K}$.

\end{abstract}

\section{Introduction}
  The lightest scalar mesons
$a_0(980)$ and $f_0(980)$, discovered more than thirty years ago,
became the hard problem for the naive quark-antiquark ($q\bar q$)
model from the outset. Really, on the one hand the almost exact
degeneration of the masses of the isovector $a_0(980)$ and
isoscalar $f_0(980)$ states revealed seemingly the structure
similar to the structure of the vector $\rho$ and $\omega$ mesons
, and on the other hand the strong coupling of $f_0(980)$ with the
$K\bar K$ channel pointed unambiguously to a considerable part of
the  strange quark pair $s\bar s$ in the wave function of
$f_0(980)$. It was noted late in the 1970s that in the MIT bag
model there are light four-quark scalar states and suggested that
$a_0(980)$ and $f_0(980)$ might be these states \cite{jaffe}. From
that time $a_0(980)$ and $f_0(980)$ resonances came into beloved
children of the light quark spectroscopy, see, for example, Refs.
\cite{montanet,achasov-84,conf}.

Ten years later there was proposed in Ref.\cite{achasov-89} to
study radiative $\phi$ decays $\phi\to
a_0\gamma\to\eta\pi^0\gamma$ and $\phi\to
f_0\gamma\to\pi^0\pi^0\gamma$ to solve the puzzle of the lightest
scalar mesons. Over the next ten years before the experiments
(1998), this question was examined from different points of view
\cite{bramon,achasov-97,achasov-97a,achasov-97b,achasov-98}.

Now these decays have been studied not only theoretically but also
experimentally. The first measurements have been reported by the
SND \cite{snd-1,snd-2,snd-fit,snd-ivan} and CMD-2 \cite{cmd}
Collaborations which obtain the following branching ratios
$$Br(\phi\to\gamma\pi^0\eta)=(8.8\pm 1.4\pm0.9)\times 10^{-5}\
\mbox{\cite{snd-fit}},$$
 $$Br(\phi\to\gamma\pi^0\pi^0)=
(12.21\pm0.98\pm0.61)\times 10^{-5}\ \mbox{\cite{snd-ivan}}$$
 $$Br(\phi\to\gamma\pi^0\eta)=(9.0\pm 2.4\pm 1.0)\times 10^{-5},$$
$$Br(\phi\to\gamma\pi^0\pi^0)=(9.2\pm0.8\pm0.6)\times 10^{-5}\
\mbox{\cite{cmd}}.$$
 More recently the KLOE Collaboration has measured
\cite{publ,pi0publ} $$
Br(\phi\to\gamma\pi^0\eta)=(8.51\pm0.51\pm0.57)\times 10^{-5}\
\mbox{in}\  \eta\to\gamma\gamma\ \mbox{\cite{publ}},$$
$$Br(\phi\to\gamma\pi^0\eta)=(7.96\pm0.60\pm0.40)\times 10^{-5}\
\mbox{in}\ \eta\to\pi^+\pi^-\pi^0\ \mbox{\cite{publ}},$$
$$Br(\phi\to\gamma\pi^0\pi^0)= (10.9\pm0.3\pm0.5)\times 10^{-5}\
\mbox{\cite{pi0publ}},$$ in agreement with the Novosibirsk data
\cite{snd-fit,snd-ivan,cmd} but with a considerably smaller error.

In this work we present the new analysis of the recent KLOE data
on the $\phi\to\eta\pi^0\gamma$ decay \cite{publ,kloe}. In
contradistinction to \cite{publ}, we

1) treat the $a_0$ mass $m_{a_0}$ as a free parameter of the fit;

2) fit the phase $\delta $ of the interference between $\phi\to
a_0\gamma\to\eta\pi^0\gamma$ (signal) and
$\phi\to\rho^0\pi^0\to\eta \pi^0\gamma$ (background) reactions;

3) use new more precise experimental values of the input
parameters.

All formulas for the $\phi\to(a_0\gamma+\rho^0\pi^0)\to\eta
\pi^0\gamma$ reaction taking the background into account are shown
in Sec.\ref{sf}. The results of the 4 different fits are presented
in Sec.\ref{sr}. A brief summary is given in Sec.\ref{sc}.

\section{The formalism of the \lowercase{$\phi\to
a_0\gamma\to\eta\pi^0\gamma$} and
\lowercase{$\phi\to\rho^0\pi^0\to\eta \pi^0\gamma$} reactions
 }
\label{sf} In Ref.\cite{a0f0} was shown that the process $\phi\to
a_0\gamma \to\eta\pi^0\gamma$ dominates in the $\phi\to
\eta\pi^0\gamma$ decay (see also \cite{achasov-89,achasov-97},
where it was predicted in four-quark model). This was confirmed in
\cite{publ,kloe}. Nevertheless, the main background process
$\phi\to\rho\pi^0\to\eta\pi^0\gamma$ should be taken into account
also (see \cite{a0f0,publ}).

The amplitude of the background process
$\phi(p)\to\pi^0\rho^0\to\gamma(q)\pi^0(k_1)\eta(k_2)$ is
\cite{a0f0}:
\begin{equation}
M_B=\frac{g_{\phi\rho\pi}g_{\rho\eta\gamma}}{D_{\rho}(p-k_1)}
\phi_{\alpha}k_{1\mu}p_{\nu}\epsilon_{\delta}(p-k_1)_{\omega}q_{\epsilon}
\epsilon_{\alpha\beta\mu\nu}\epsilon_{\beta\delta\omega\epsilon}.
\end{equation}

According to the one-loop mechanism of the decay $\phi\to
K^+K^-\to\gamma a_0$, suggested in Ref.\cite{achasov-89}, the
amplitude of the signal $\phi\to\gamma a_0\to\gamma\pi^0\eta$ has
the form:

\begin{equation}
M_a=g(m)\frac{g_{a_0K^+K^-}g_{a_0\pi\eta}}{D_{a_0}(m)}\bigg(
(\phi\epsilon)- \frac{(\phi q)(\epsilon p)}{(pq)} \bigg)
\label{a0signal}\,,
\end{equation}
where  $m^2=(k_1+k_2)^2$, $\phi_{\alpha}$ and $\epsilon_{\mu}$ are
the polarization vectors of $\phi$ meson and photon, the forms of
$g_R(m)$ and $g(m)=g_R(m)/g_{RK^+K^-}$ everywhere over the $m$
region are in Refs. \cite{achasov-89} and \cite{achasov-01a}
respectively:

 For $m<2m_{K^+}$

\begin{eqnarray}
&&g(m)=\frac{e}{2(2\pi)^2}g_{\phi K^+K^-}\Biggl\{
1+\frac{1-\rho^2(m^2)}{\rho^2(m^2_{\phi})-\rho^2(m^2)}\times\nonumber\\
&&\Biggl[2|\rho(m^2)|\arctan\frac{1}{|\rho(m^2)|}
-\rho(m^2_{\phi})\lambda(m^2_{\phi})+i\pi\rho(m^2_{\phi})-\nonumber\\
&&-(1-\rho^2(m^2_{\phi}))\Biggl(\frac{1}{4}(\pi+
i\lambda(m^2_{\phi}))^2- \nonumber\\
&&-\Biggl(\arctan\frac{1}{|\rho(m^2)|}\Biggr)^2
\Biggr)\Biggr]\Biggr\},
\end{eqnarray}
where
\begin{equation}
\rho(m^2)=\sqrt{1-\frac{4m_{K^+}^2}{m^2}}\,\,;\qquad
\lambda(m^2)=\ln\frac{1+\rho(m^2)}{1-\rho(m^2)}\,\,;\qquad
\frac{e^2}{4\pi}=\alpha=\frac{1}{137}\,\,.
\end{equation}

 For $m\geq 2m_{K^+}$
\begin{eqnarray}
&&g(m)=\frac{e}{2(2\pi)^2}g_{\phi K^+K^-}\Biggl\{
1+\frac{1-\rho^2(m^2)}{\rho^2(m^2_{\phi})-\rho^2(m^2)}\times\nonumber\\
&&\times\Biggl[\rho(m^2)(\lambda(m^2)-i\pi)-
\rho(m^2_{\phi})(\lambda(m^2_{\phi})-i\pi)-\nonumber\\
&&\frac{1}{4}(1-\rho^2(m^2_{\phi}))
\Biggl((\pi+i\lambda(m^2_{\phi}))^2-
(\pi+i\lambda(m^2))^2\Biggr)\Biggr]\Biggr\}.
\end{eqnarray}

  The mass spectrum is
\begin{equation}
\frac{d\Gamma(\phi\to\gamma\pi^0\eta,m)}{dm}=\frac{d\Gamma_{a_0}(m)}{dm}+
\frac{d\Gamma_{back}(m)}{dm}+ \frac{d\Gamma_{int}(m)}{dm}\,,
\end{equation}
where the mass spectrum for the signal is
\begin{equation}
\frac{d\Gamma_{a_0}(m)}{dm}=\frac{2}{\pi}\frac{m^2\Gamma(\phi\to\gamma
a_0,m)\Gamma(a_0\to\pi^0\eta,m)}{|D_{a_0}(m)|^2}=
\frac{2|g(m)|^2p_{\eta\pi}(m_{\phi}^2-m^2)}
{3(4\pi)^3m_{\phi}^3}\bigg|\frac{g_{a_0K^+K^-}g_{a_0\pi\eta}}{D_{a_0}(m)}\bigg|^2\,.
\label{spectruma0}
\end{equation}

The mass spectrum for the background process
$\phi\to\pi^0\rho\to\gamma\pi^0\eta$  is \cite{a0f0}:

\begin{equation}
\frac{d\Gamma_{back}(m)}{dm}=\frac{(m_{\phi}^2-m^2)p_{\pi\eta}
}{128\pi^3m_{\phi}^3}\int_{-1}^{1}dxA_{back}(m,x)\,,
\end{equation}
where
\begin{eqnarray}
&&A_{back}(m,x)=\frac{1}{3}\sum|M_B|^2= \nonumber \\
&&=\frac{1}{24}(m_{\eta}^4m_{\pi}^4+2m^2m_{\eta}^2m_{\pi}^2
\tilde{m_{\rho}}^2-2m_{\eta}^4m_{\pi}^2\tilde{m_{\rho}}^2-2m_{\eta}^2m_{\pi}^4
\tilde{m_{\rho}}^2+\nonumber \\ &&2m^4\tilde{m_{\rho}}^4-
2m^2m_{\eta}^2\tilde{m_{\rho}}^4+m_{\eta}^4\tilde{m_{\rho}}^4
-2m^2m_{\pi}^2\tilde{m_{\rho}}^4+4m_{\eta}^2m_{\pi}^2\tilde{m_{\rho}}^4
+m_{\pi}^4\tilde{m_{\rho}}^4+\nonumber \\
&&2m^2\tilde{m_{\rho}}^6-
2m_{\eta}^2\tilde{m_{\rho}}^6-2m_{\pi}^2\tilde{m_{\rho}}^6+
\tilde{m_{\rho}}^8-2m_{\eta}^4m_{\pi}^2m_{\phi}^2-
2m^2m_{\eta}^2m_{\phi}^2\tilde{m_{\rho}}^2+\nonumber \\
&&2m_{\eta}^2m_{\pi}^2m_{\phi}^2\tilde{m_{\rho}}^2-
2m^2m_{\phi}^2\tilde{m_{\rho}}^4+
2m_{\eta}^2m_{\phi}^2\tilde{m_{\rho}}^4-
2m_{\phi}^2\tilde{m_{\rho}}^6+ m_{\eta}^4m_{\phi}^4+
m_{\phi}^4\tilde{m_{\rho}}^4)\times\nonumber \\
&&\bigg|\frac{g_{\phi\rho\pi}g_{\rho\eta\gamma}}
{D_{\rho}(\tilde{m_{\rho}})}\bigg|^2\,, \label{Aback}
\end{eqnarray}
and
\begin{eqnarray}
&&\tilde{m_{\rho}}^2=m_{\eta}^2+\frac{(m^2+m_{\eta}^2-m_{\pi}^2)(m_{\phi}^2-
m^2)}{2m^2}-\frac{(m_{\phi}^2-m^2)x}{m}p_{\pi\eta}\nonumber \\
&&p_{\pi\eta}=\frac{\sqrt{(m^2-(m_{\eta}-m_{\pi})^2)
(m^2-(m_{\eta}+m_{\pi})^2)}}{2m}\,.
\end{eqnarray}

Note that there is a misprint in Eq.(6) of Ref.\cite{a0f0}, which
describes $A_{back}(m,x)$: the 7th term in the brackets
"$+2m_{\eta}^4\tilde{m_{\rho}}^4$" should be replaced by
"$+m_{\eta}^4\tilde{m_{\rho}}^4$", as above in Eq.(\ref{Aback}).
Emphasize that all evaluations in Ref.\cite{a0f0} were done with
the correct formula.

The term of the interference between the signal and the background
processes is written in the following way:

\begin{equation}
\frac{d\Gamma_{int}(m)}{dm}=\frac{(m_{\phi}^2-m^2)p_{\pi\eta}
}{128\pi^3m_{\phi}^3} \int_{-1}^{1}dxA_{int}(m,x)\,,
\end{equation}
where
\begin{eqnarray}
&&A_{int}(m,x)=\frac{2}{3}Re\sum M_aM_B^*=
\frac{1}{3}\left((m^2-m_{\phi}^2)\tilde{m_{\rho}}^2+
\frac{m_{\phi}^2(\tilde{m_{\rho}}^2-m_{\eta}^2)^2}{m_{\phi}^2-m^2}\right)
\times\nonumber\\ &&Re\{\frac{e^{i\delta
}g(m)g_{a_0K^+K^-}g_{a_0\pi\eta}g_{\phi\rho\pi}g_{\rho\eta\gamma}}
{D^*_{\rho}(\tilde{m_{\rho}})D_{a_0}(m)}\}\,.
\end{eqnarray}

Note that the phase $\delta$ isn't taken into account in
\cite{a0f0}. The inverse propagator of the scalar meson R ($a_0$
in our case), is presented in Refs.
\cite{achasov-80,z_phys,achasov-89,achasov-97}:

\begin{equation}
\label{propagator} D_R(m)=m_R^2-m^2+\sum_{ab}[Re
\Pi_R^{ab}(m_R^2)-\Pi_R^{ab}(m^2)],
\end{equation}
where $\sum_{ab}[Re \Pi_R^{ab}(m_R^2)-
\Pi_R^{ab}(m^2)]=Re\Pi_R(m_R^2)- \Pi_R(m^2)$ takes into account
the finite width corrections of the resonance which are the one
loop contribution to the self-energy of the $R$ resonance from the
two-particle intermediate  $ab$ states.

For the pseudoscalar $ab$ mesons and $m_a\geq m_b,\ m\geq m_+$ one
has
\cite{achasov-80,z_phys,achasov-97b,achasov-84,achasov-95}\footnote{
Note that in Ref.\cite{achasov-80} $\Pi_R^{ab}(m^2)$ differs by a
real constant from those determined in other enumerated works in
the case of $m_a\neq m_b$, but obviously it has no effect on
Eq.(\ref{propagator}). }:

\begin{eqnarray}
\label{polarisator}
&&\Pi^{ab}_R(m^2)=\frac{g^2_{Rab}}{16\pi}\left[\frac{m_+m_-}{\pi
m^2}\ln \frac{m_b}{m_a}+\right.\nonumber\\
&&\left.+\rho_{ab}\left(i+\frac{1}{\pi}\ln\frac{\sqrt{m^2-m_-^2}-
\sqrt{m^2-m_+^2}}{\sqrt{m^2-m_-^2}+\sqrt{m^2-m_+^2}}\right)\right]
\end{eqnarray}
For $m_-\leq m<m_+$
\begin{eqnarray}
&&\Pi^{ab}_{R}(m^2)=\frac{g^2_{Rab}}{16\pi}\left[\frac{m_+m_-}{\pi
m^2}\ln \frac{m_b}{m_a}-|\rho_{ab}(m)|+\right.\nonumber\\
&&\left.+\frac{2}{\pi}|\rho_{ab}(m)
|\arctan\frac{\sqrt{m_+^2-m^2}}{\sqrt{m^2-m_-^2}}\right].
\end{eqnarray}
For $m<m_-$
\begin{eqnarray}
&&\Pi^{ab}_{R}(m^2)=\frac{g^2_{Rab}}{16\pi}\left[\frac{m_+m_-}{\pi
m^2}\ln \frac{m_b}{m_a}-\right.\nonumber\\
&&\left.-\frac{1}{\pi}\rho_{ab}(m)\ln\frac{\sqrt{m_+^2-m^2}-
\sqrt{m_-^2-m^2}}{\sqrt{m_+^2-m^2}+\sqrt{m_-^2-m^2}}\right].
\end{eqnarray}
and
\begin{equation}
\label{rho-ab}
\rho_{ab}(m)=\sqrt{(1-\frac{m_+^2}{m^2})(1-\frac{m_-^2}{m^2})}\,\,,\qquad
m_{\pm}=m_a\pm m_b
\end{equation}
The constants  $g_{Rab}$ are related to the width
\begin{equation}
\Gamma(R\to ab,m)=\frac{g_{Rab}^2}{16\pi m}\rho_{ab}(m).
\label{f0pipi}
\end{equation}

In our case we take into account intermediate states
$ab=\eta\pi^0,\ K\bar{K}$ and $\eta '\pi^0$:

\begin{equation}
\Pi_{a_0}=\Pi_{a_0}^{\eta\pi^0}+\Pi_{a_0}^{K^+K^-}+
\Pi_{a_0}^{K^0\bar{K^0}}+\Pi_{a_0}^{\eta' \pi^0},
\end{equation}

\noindent $g_{a_0K^+K^-}=-g_{a_0K^0\bar{K^0}}$. Note that the
$\eta '\pi^0$ contribution is of small importance due to the high
threshold. Even fitting with $|g_{a_0\eta '\pi^0}| =0$ changes the
results less than 10\% of their errors. We set $|g_{a_0\eta
'\pi^0}| =|1.13\,g_{a_0K^+K^-}|$ according to the four-quark
model, see \cite{achasov-89}, but this is practically the same as
the 2-quark model prediction $|g_{a_0\eta '\pi^0}|
=|1.2\,g_{a_0K^+K^-}|$, see \cite{achasov-89}.

The inverse propagator of the $\rho$ meson has the
following expression
\begin{equation}
D_{\rho}(m)=m_{\rho}^2-m^2-im^2\frac{g^2_{\rho\pi\pi}}{48\pi}
\bigg(1-\frac{4m_{\pi}^2}{m^2}\bigg)^{3/2}\,.
\end{equation}

The coupling constants $g_{\phi K^+K^-}=4.376\pm 0.074$ and
$g_{\phi\rho\pi}=0.814\pm0.018$ GeV$^{-1}$ are taken from the new
most precise measurement Ref.\cite{sndphi}. Note that in Ref.
\cite{publ,a0f0} the value $g_{\phi K^+K^-}=4.59$ was obtained
using the \cite{pdg} data. The coupling constant
$g_{\rho\eta\gamma}=0.56\pm 0.05$ GeV$^{-1}$ is obtained from the
data of Ref.\cite{pdg-2002} with the help of the expression
\begin{equation}
\Gamma(\rho\to\eta\gamma)=\frac{g_{\rho\eta\gamma}^2}{96\pi
m_{\rho}^3} (m_{\rho}^2-m_{\eta}^2)^3.
\end{equation}
\begin{figure} \centerline{
\epsfxsize=12 cm \epsfysize=8cm \epsfbox{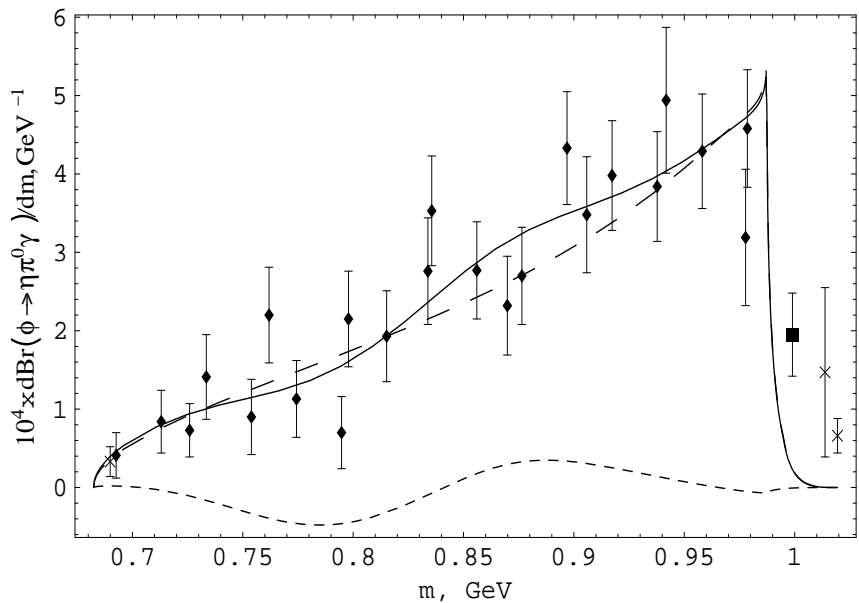}}
 \caption{The comparison of fit 1 (solid line) with the KLOE data
 (points). The signal contribution and the interference term are
shown with the dashed line and the dotted line. Cross points are
omitted in fitting. The box point (0.999 GeV) is omitted in fits 3
and 4. } \label{fig1}
\end{figure}

\section{Results}
\label{sr} The KLOE data on the $\phi\to\eta\pi^0\gamma$ decay may
be found in Table 5 of Ref.\cite{kloe} (see also Fig.\ref{fig1}).
Note that as in Refs.\cite{publ,kloe}, we don't fit 1st,10th and
27th points of this table (cross points in the Fig. \ref{fig1}).
Emphasize that the 10th (1.014 GeV) and 27th (1.019 GeV) points
are obvious artifacts because the mass spectrum behaviour on the
right slope of the resonance has the form the (photon energy)$^3$
according to gauge invariance.

In the experiment the whole mass region
($m_{\eta}+m_{\pi^0},m_{\phi}$) is divided into some number of
bins. Experimenters measure the average value ${\bar{B}_i}$ ("i"
is the number of bin) of $dBr(\phi\to\eta\pi^0\gamma)/dm$ around
each i-th bin:

\begin{equation}
\bar{B}_i=\frac{1}{m_{i+1}-m_i }\int ^{m_{i+1}}_{m_i}
dBr(\phi\to\eta\pi^0\gamma)/dm,
\end{equation}

In this case one should define $\chi^2$ function as:

\begin{equation}
\chi^2=\sum_i \frac{(\bar{B}_i^{th}-\bar{B}_i^{exp})^2}{\sigma
_i^2},
\end{equation}

\noindent where $\bar{B}_i^{exp}$ are the experimental results,
$\sigma _i$ are the experimental errors, and

$$\bar{B}_i^{th}=\frac{1}{m_{i+1}-m_i }\int ^{m_{i+1}}_{m_i}
dBr^{th}(\phi\to\eta\pi^0\gamma)/dm$$

($dBr^{th}(\phi\to\eta\pi^0\gamma)/dm$ is the theoretical curve).

The free parameters of the fit are $m_{a_0},
g_{a_0K^+K^-}^2/4\pi$, the phase $\delta $ (we assume it is
constant) and the ratio $g_{a_0\eta\pi}/g_{a_0K^+K^-}$. The
results are shown in Table 1 (fit 1) \footnote{Note that fitting
without averaging the theoretical curve $\bigg($changing
$\bar{B}_i^{th}\to dBr^{th}(\phi\to\eta\pi^0\gamma)/dm\bigg|
_{m=(m_{i+1}+m_i)/2}\bigg)$ results to worse
$\chi^2/n.d.f.=28.8/20$. The results in this case are consistent
within errors with those obtained with averaging the theoretical
curve.}.

The quality of the fit is good. The phase $\delta $ is consistent
with zero, so we make a fit with $\delta =0$ (fit 2 in Table 1).

To check the correctness of treating the phase $\delta $ as a
constant, we have done a fit with $\delta$ taken in the form
$\delta(m)=bp_{\eta\pi}(m)$ (the phase of the elastic background
in $\eta\pi^0$ scattering may have such behaviour), and found out
that the constant b=$(2.8\pm 3.2)$ GeV$^{-1}$ is also consistent
with zero. Change of the other values is not principal.

\begin{center}
\begin{tabular}{|c|c|c|c|c|c|}
\multicolumn{4}{c}{Table 1.} \\ \hline

fit & $m_{a_0}$, MeV & $\frac{g_{a_0K^+K^-}^2}{4\pi}$, GeV$^2$ &
$g_{a_0\eta\pi}/g_{a_0K^+K^-}$ & $\delta $, $^{\circ}$ &
$\chi^2/n.d.f.$ \\ \hline
 1 & \hspace{1mm} $1003^{+32}_{-13}$ \hspace{1mm} &
$0.82^{+0.81}_{-0.27}$ & $1.06^{+0.20}_{-0.27}$ & $27 \pm 29$ &
24.2/20
\\ \hline 2 & $995^{+22}_{-8}$ & $ \hspace{1mm} 0.65^{+0.42}_{-0.18}
\hspace{1mm}$ & $1.17^{+0.17}_{-0.24}$ & 0 & 25.2/21 \\ \hline
 3 & \hspace{1mm} $994^{+22}_{-8}$ \hspace{1mm} &
$0.62^{+0.4}_{-0.17}$ & $1.21^{+0.17}_{-0.24}$ & $21 \pm 30$  &
16.3/19
\\ \hline
 4 & \hspace{1mm} $992^{+14}_{-7}$ \hspace{1mm} &
$0.55^{+0.27}_{-0.13}$ & $1.26^{+0.16}_{-0.2}$ & $ 0 $ & 16.9/20
\\ \hline
\end{tabular}
\end{center}

Since the discrepancy between fits and the experimental point
number 26 (0.999 GeV) in the Table 5 of Ref.\cite{kloe} (the box
point in Fig.1) is about 3 standard deviations (i.e. this point
may be an artifact also), we make another fit without this point
(fit 3). The phase $\delta$ is again consistent with zero, so we
make a fit without it (fit 4).

In Table 2 we present the results on the total branching ratio
$\mbox{Br}(\phi\to(a_0\gamma+\rho\pi^0)\to\eta\pi^0\gamma)$, the
signal contribution $\mbox{Br}(\phi\to
a_0\gamma\to\eta\pi^0\gamma)$, $\Gamma_{a_0\eta\pi^0}\equiv
\Gamma(a_0\to\eta\pi^0,m_{a_0})=g_{a_0\eta\pi}^2 \rho_{\eta\pi^0 }
(m_{a_0})/(16\pi m_{a_0})$ and the ratio
$R=g^2_{f_0K^+K^-}/g^2_{a_0K^+K^-}$. The last is obtained using
the Ref.\cite{pi0publ} value $g^2_{f_0K^+K^-}/(4\pi)=2.79\pm0.12$
GeV$^2$. The branching ratio of the background
$\mbox{Br}(\phi\to\rho\pi^0\to\eta\pi^0\gamma)$ accounts for
$(0.5\pm 0.1) \times 10^{-5}$.

\begin{center}
\begin{tabular}{|c|c|c|c|c|}
\multicolumn{4}{c}{Table 2.} \\ \hline

fit & $\Gamma_{a_0\eta\pi^0}$, MeV & R & $10^5\times
\mbox{Br}(\phi\to (a_0\gamma + \rho\pi ^0) \to\eta\pi ^0\gamma)$ &
$10^5\times \mbox{Br}(\phi\to a_0\gamma\to\eta\pi^0\gamma)$ \\
\hline
 1 & $153^{+22}_{-17}$ & $3.4\pm 1.7$ & $7.6\pm 0.4$ & $7.3\pm 0.4$ \\ \hline
2 & $148^{+17}_{-15}$ & $4.3\pm 1.7$ & $7.6\pm 0.4$ & $7.1\pm 0.4$
\\ \hline
 3 & $149^{+19}_{-16}$ & $4.5\pm 1.7$ & $7.6\pm 0.4$ & $7.2\pm 0.4$ \\ \hline
 4 &$146^{+17}_{-15}$& $5.0\pm 1.6$ & $7.6\pm 0.4$ & $7.1\pm 0.4$ \\ \hline
\end{tabular}

\end{center}

\section{Conclusion}
\label{sc} Note that the obtained value of the ratio
$g_{a_0\eta\pi}/g_{a_0K^+K^-}$ doesn't contradict the first
predictions based on the four-quark model of the $a_0$:
$g_{a_0\eta\pi}/g_{a_0K^+K^-}\approx 0.85$
\cite{achasov-89}\footnote{Note that the prediction
$g_{a_0\eta\pi}/g_{a_0K^+K^-}\approx 0.93$, made in \cite{jaffe},
was corrected in \cite{mistake} }. But even if
$g_{a_0\eta\pi}/g_{a_0K^+K^-}$ deviates from $0.85$, there is no
tragedy, because those variant of the four-quark model is rather
rough, it is considered as a guide.

For all fits the obtained value R differs from the \cite{publ}
value $R=7.0\pm0.7$. So the conclusion that the constant
$g^2_{a_0K^+K^-}/(4\pi)$ is small, obtained in \cite{publ,kloe}
($g^2_{a_0K^+K^-}/(4\pi)=0.4\pm 0.04$ GeV$^2$), is the result of
the parameters restrictions, especially fixing $m_{a_0}$ at the
PDG-2000 value 984.8 MeV. Note that a high $a_0$ mass is also
needed to describe $\gamma\gamma\to\eta\pi^0$ experiment, see
\cite{lmas}.

There should be no confusion due to the large $a_0$ width. In the
peripheral production of the $a_0$ (for example, in the reaction
$\pi^-p\to\eta\pi^0 n$) the mass spectrum is given by the relation

\begin{equation}
\frac{dN_{\eta\pi^0}}{dm}\sim S_{per}(m)=\frac{2m^2}{\pi}
\frac{\Gamma(a_0\to\eta\pi^0,m)}{|D_{a_0}(m)|^2}.
\end{equation}

\begin{figure}
\centerline{ \epsfxsize=12 cm \epsfysize=8cm
\epsfbox{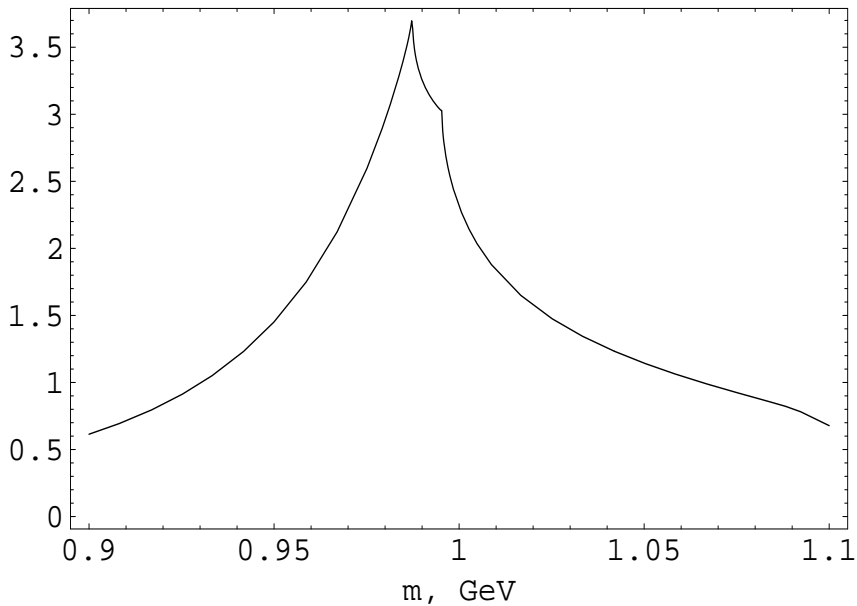}}
 \caption{ Plot of the function $S_{per}(m)$, GeV$^{-1}$, for fit 1 results.}
\label{fig2}
\end{figure}

The effective (visible) width of this distribution is much less
then the nominal width $\Gamma_{a_0\eta\pi^0}$. For example, for
the fit 1 results (Table 1) the effective width is $\sim 50$ MeV,
see Fig. \ref{fig2}.

As it is noted in Ref.\cite{our}, there is no tragedy with the
relation between branching ratios of $a_0$ and $f_0$ production in
$\phi$ radiative decays. The early predictions \cite{achasov-89}
are based on the one-loop mechanism $\phi\to K^+K^-\to
a_0\gamma\to\eta\pi^0\gamma$ and $\phi\to K^+K^-\to
f_0\gamma\to\pi\pi\gamma$ at $m_{a_0}=980$ MeV, $m_{f_0}=975$ MeV
and $g_{a_0K^+K^-}=g_{f_0K^+K^-}$\footnote{Emphasize that the
isotopic invariance doesn't require
$g_{a_0K^+K^-}=g_{f_0K^+K^-}$.}, that leads to $\mbox{Br}(\phi\to
a_0\gamma\to\eta\pi^0\gamma)\approx \mbox{Br}(\phi\to
f_0\gamma\to\pi\pi\gamma)$. But it is shown in Ref.
\cite{achasov-97}, that the relation between branching ratios of
$a_0$ and $f_0$ production in $\phi$ radiative decays essentially
depends on a $a_0-f_0$ mass splitting. This strong mass dependence
is the result of gauge invariance, the (photon energy)$^3$ law on
the right slope of the resonance. Our present analysis confirms
this conclusion. Note that a noticeable deviation from the naive
four-quark model equality $g_{a_0K^+K^-}=g_{f_0K^+K^-}$ is not
crucial. What is more important is the mechanism of the production
of the $a_0$ and $f_0$ through the charged kaon loop, i.e. the
four-quark transition. As is shown in Ref.\cite{conf}, this gives
strong evidence in favor of the four-quark model of the
$a_0(f_0)$.

Note that the constant $g^2_{f_0K^+K^-}/(4\pi)$ also can differ a
lot from those obtained in \cite{pi0publ}. The point is that the
extraction of this constant is very model dependent. For example,
fitting with taking into account the mixing of the resonances can
decrease the value of $g^2_{f_0K^+K^-}/(4\pi)$ considerably. For
instance, fitting the data of \cite{snd-ivan} without mixing one
has $g^2_{f_0K^+K^-}/(4\pi)=2.47^{+0.37}_{-0.51}$ GeV$^2$
\cite{snd-ivan}, while fitting with taking the mixing into account
gives $g^2_{f_0K^+K^-}/(4\pi)=1.29\pm 0.017$ GeV$^2$ \cite{a0f0}.
Note also that in \cite{pi0publ} the phase $\delta_B$ of the
background is taken from the work \cite{a0f0}, where it is
obtained by the simultaneous fitting the $m_{\pi^0\pi^0}$ spectrum
and the phase $\delta_0$ of the $\pi\pi$ scattering, taking into
account the mixing of the resonances. In \cite{pi0publ} the mixing
isn't taken into account, so additional phase dealing with it is
omitted.

\section{Acknowledgements}
We thank C. Bini very much for providing the useful information,
discussions and kind contacts. This work was supported in part by
RFBR, Grant No 02-02-16061.

\end{document}